\documentclass[aps,superscriptaddress]{revtex4}
\usepackage{graphicx}

\begin{document}
%\firstpage{1}
 
\title{Clustering by soft-constraint 
affinity propagation:\\
Applications to gene-expression data} 
\author{Michele Leone\,$^{\rm a}$, 
Sumedha\,$^{\rm a}$, Martin Weigt\,$^{\rm a}$}
\address{$^{\rm a}$Institute for Scientific Interchange, Viale 
Settimio Severo 65, Villa Gualino, I-10133 Torino, Italy.}
%\maketitle
\begin{abstract}
{\it Motivation:}
Similarity-measure based clustering is a crucial problem appearing 
throughout scientific data analysis. Recently, a powerful new algorithm 
called Affinity Propagation (AP) based on message-passing techniques was 
proposed by  Frey and Dueck \cite{Frey07}. In AP, each
cluster is identified by a common exemplar all other data points of the 
same cluster refer to, and exemplars have to refer to themselves.
Albeit its proved power, AP in its present form suffers from a number of 
drawbacks. The hard constraint of having exactly one exemplar per cluster 
restricts AP to classes of regularly shaped 
clusters, and leads to suboptimal performance, {\it e.g.}, in analyzing gene
expression data.

{\it Results:}
This limitation can be overcome by relaxing the AP hard constraints. 
A new parameter controls the importance of the
constraints compared to the aim of maximizing the overall similarity, and
allows to interpolate between the simple case where each data point selects
its closest neighbor as an exemplar and the original AP. The
resulting soft-constraint affinity propagation (SCAP) becomes more 
informative, accurate and leads to more stable clustering. Even though a new 
{\it a priori} free-parameter is introduced, the overall dependence of the 
algorithm on external tuning is reduced, as robustness is increased and an 
optimal strategy for parameter selection emerges more naturally.
SCAP is tested on biological benchmark data, including in particular
microarray data related to various cancer types. We show that the algorithm
efficiently unveils the hierarchical cluster structure present in the data
sets. Further on, it allows to extract sparse gene expression signatures
for each cluster.

%\section{Availability:}
%Text  Text Text Text Text Text Text Text
 
{\it Contact:}{leone,sumedha,weigt@isi.it}
\end{abstract}  

\maketitle

\section{Introduction}
Clustering based on a measure of similarity is a crucial problem which appears 
throughout scientific data analysis. For an overview see \cite{Jain99}.
Recently, a powerful algorithm called Affinity Propagation (AP)  based on {\it message
passing} was proposed by  Frey and Dueck \cite{Frey07}. As reported impressively in
the original publication, this algorithm achieves a considerable improvement over 
standard clustering methods like K-means \cite{MacQueen67}, spectral 
clustering \cite{Shi00} and super-paramagnetic clustering \cite{blatt96,Shental}.

Based on an {\sl ad hoc} pairwise similarity function between data points,
AP seeks to identify each cluster by one of its elements, the so-called {\it
exemplar}. Each point in the cluster refers to this exemplar, each exemplar
is required to refer to itself as a self-exemplar. This hard constraint forces
clusters to appear as stars of radius one: There is only one central 
node, and all other nodes are directly connected to it. Subject to this constraint,
AP seeks at maximizing the overall similarity of all data points to their exemplars. 
The solution to this hard combinatorial task is approximated following the ideas of 
belief-propagation \cite{Yedidia05,Kschischang01}. One of the important points 
of AP is its computational efficiency: whereas a naive implementation of belief 
propagation 
for $N$ data points leads to $O(N^3)$ messages which have to be determined 
self-consistently, the elegant formulation of Frey and Dueck allows to work
with $O(N^2)$ messages only. Therefore the algorithm is feasible even in 
the presence of very large data sets.

Albeit its impressive power in a wide range of applications \cite{Frey07}, AP in 
its present form suffers from a number of drawbacks. The most important ones
related to the present work are:
%\footnote{Other limitations may lie in the choice of a non optimal similarity 
%function, or in the pairwise nature of the interaction that does not take into 
%account that the similarity between two data points could be dependent on the 
%presence or absence of other points in the data set. This would result in an 
%effective many-body similarity function. However, these effects are not easily
%measured in raw biological data and/or imply some degree of prior knowledge of 
%the sought clustering solution, which is usually not available in unsupervised 
%clustering schemes like the one implemented via the AP algorithm.}
\begin{itemize}
\item The hard constraint in AP relies strongly on {\it cluster-shape regularity}:
Elongated or irregular multi-dimensional data might have more than one
simple cluster center. AP may force division
of single clusters into separate ones.
\item Since all data points in a cluster must point to the same exemplar,
all information about both the {\it internal structure} and the
{\it hierarchical merging/dissociation} of clusters is lost.
\item AP has {\it robustness limitations}: A small perturbation of 
similarities may influence the choice of one or few exemplars, and the
hard constraint may trigger an avalanche leading to a different partitioning
of the data set into clusters. This point is particularly important in the
presence of noise in the data as, {\it e.g.}, in microarray measurements. 
\item AP forces each exemplar to point to itself. A relaxation of the hard 
constraint may allow for cluster
structures including {\it second- or higher-order pointing processes}.
\end{itemize}

These problems may be solved by modifying the original optimization task of AP. 
As a first step we relax the hard constraint by introducing a finite penalty term 
for each constraint
violation. This softening can be chosen in a way that the computational
complexity of the algorithm remains unchanged, but its performance on biological
test sets is improved considerably. Moreover, relaxing the constraint helps in 
gaining valuable insight into the hierarchical structure of the clustering, 
increasing result robustness at the same time. By tuning the cluster number we 
see the merging of two clusters into a single one, or the dissociation of single 
clusters into two separated ones.

%In this work, we first introduce the modifications leading to SCAP. Then
%we apply it to various biological test data sets. In the methods section we provide
%a detailed derivation of the algorithm starting from the original optimization
%task.

\section{The Algorithm}

Given a set $D = \{ x_\mu \}_{\mu = 1}^N$ of $N$ data points, 
the original algorithm of Frey and Dueck takes as input a collection of
real valued  similarities $S(\mu,\nu)$ between the pairs 
$x_\mu,x_\nu,\ \mu,\nu\in\{1,...,N\}$. The choice of the similarity measure is 
heuristic, it depends on the nature of data to be
clustered. In the case of high-dimensional data as present in gene-expression
analysis, similarity may be measured by the Pearson correlation coefficient
or the negative pairwise Euclidean distance. However, for the algorithm
described below it is not even necessary that the similarities are symmetric.

The algorithm searches for a mapping ${\bf c}: \{1,...,N\} \mapsto \{1,...,N\}$ which
maps each data point $\mu$ to its exemplar $c_\mu$ which itself is a data point.
This mapping shall minimize the cost function (or energy)
$E_1[{\bf c}] = -\sum_{\mu=1}^N S(\mu,c_\mu)$
which equals minus the overall similarity of the data points to their exemplars.
In the original AP algorithm ${\bf c}$ is restricted by $N$ hard constraints: Whenever
a data point is selected as an exemplar by another data point, it has to be its
own self-exemplar.

In this setting, we need to specify self-similarities $S(\mu,\mu)$. 
They describe the availability of data points for being self-exemplars 
(and thus to serve as a cluster center). Since all data 
points are {\it a priori} equally suitable to play such a role, one is naturally 
led to choose all self-similarities to have some common value 
$S(\mu,\mu)\equiv-\sigma,\ \mu=1,...,N$.
%\footnote{Prior biological knowledge leading to a preference of some data points to
%be exemplars may be implemented via a non-homogeneous self-similarity.}
The free model parameter $\sigma$ acts 
like a chemical potential in statistical physics, setting the prior likelihood of
the number of self exemplars (and of separated clusters consequently). For very
small value of $\sigma$, 
every data point prefers to be its own exemplar, and the number of clusters equals 
the number of data points. In the opposite extreme case of large $\sigma$, 
self-exemplars have high cost
in $E_1[{\bf c}]$. All data points are collected in one large cluster with a single
exemplar. For intermediate values $\sigma$ acts
as a tuning parameter for the cluster number which decreases monotonously with
$\sigma$. Frey and Dueck argue that, if the data set has some underlying structure, 
the correct clustering can be identified by a comparably broad range of values
of the self-similarity in which the inferred cluster structure does not change.
If data are not sparse and clusters are symmetrically shaped, then
affinity propagation works very well and  produces the correct clustering in a 
very short time.

Finding the cost minimum of $E_1[{\bf c}]$ subject to the self-exemplar constraint is 
a computationally hard task. It can be achieved exactly only for very small
systems. The central idea of AP is therefore to identify the exemplars using
message passing (belief propagation, BP) as a heuristic strategy \cite{Yedidia05}: 
Real-valued messages 
between pairs of data points are updated recursively until a stable clustering 
emerges. The original AP equations are a direct application of BP
(or, equivalently, Max-Sum \cite{Kschischang01}) to the clustering 
problem.

There are two types of messages exchanged between data points \cite{Frey07}: 
The {\it responsibility} $r(\mu,\nu)$ is sent from data point $\mu$ to candidate
exemplar $\nu$; it reflects the accumulated evidence that $\mu$ chooses $\nu$ as 
its cluster exemplar. The {\it availability} $a(\mu,\nu)$ is
sent from candidate exemplar $\nu$ to datum $\mu$; it reflects the appropriateness
for $\nu$ to be an exemplar for $\mu$ as a result of the self-exemplar constraint.
As mentioned before, the original AP imposes constraints on exemplars to be
self-exemplars. We modify the algorithm of Frey and Dueck by softening
this hard constraint. We write the constraint attached to a
given data point $\mu$ as follows, with $p \in [ 0, 1 ]$:
\begin{equation}
  \chi^{(p)}_{\mu} [{\bf c}] = \left\{
  \begin{array}{lll}
    p & {\rm if} & \exists \nu:~c_{\nu}=\mu,~c_{\mu} \ne \mu \\[0.1cm]
    1 & {\rm else.} &
  \end{array}
  \right.
  \label{constraints}
\end{equation}
The first case assigns a penalty $p$ if data point $\mu$ is chosen as an exemplar
by some other data point $\nu$, without being a self-exemplar. The hard-constraint
limit of Frey and Dueck is recovered by setting $p$ to zero. For $p=1$, 
$\chi^{(p)}_{\mu}[{\bf c}]$ becomes identically one, the minimization task of 
$E_1[{\bf c}]$ becomes unconstrained and independent for all data points, thus
each data point chooses his nearest neighbor as an exemplar. An intermediate value 
of $p$ allows to interpolate between these two extreme cases. It presents a compromise 
between the minimization of $E_1[{\bf c}]$ on one hand, and the search for compact
clusters on the other hand.

Finally we introduce a positive real-valued parameter $\beta$ weighing the 
relative importance of the cost minimization with respect to the constraints. In a
statistical-physics perspective, this parameter can be seen as a formal inverse
temperature. Its introduction allows us to define the probability of an arbitrary 
clustering ${\bf c}$ as
\begin{equation}
P[{\bf c}] = \frac{1}{Z} \exp\left( - \beta E_1[{\bf c}] \right)\
\prod_\mu \ \chi^{(p)}_{\mu} [{\bf c}]  
\label{partition}
\end{equation}
where the partition function $Z$ serves to normalize $P[{\bf c}]$. In this setting, 
both clusterings of non-optimal cost and of many violated self-exemplar constraints 
are suppressed exponentially. The task of finding high-scoring ${\bf c}$ can be 
understood as a minimization problem with the modified cost function
\begin{equation}
E_p[{\bf c}] = -\sum_{\mu=1}^{N} S(\mu,c_{\mu})- \frac 1\beta \sum_{\mu=1}^{N}
\ln\left( \chi^{(p)}_{\mu}[{\bf c}] \right)
\label{energy}
\end{equation}
AP is recovered by taking $p=0$ since any violated constraint
sets $P[{\bf c}]$ to zero in Eq.~(\ref{partition}). For general $p$, the optimal 
clustering ${\bf c}^\star$ can be determined component-wise by maximizing the marginal 
probabilities,
\begin{equation}
c^\star_\mu = {\rm argmax}_{c_\mu} P_\mu (c_\mu)
= {\rm argmax}_{c_\mu} \sum_{\{c_\nu;\nu\neq\mu\}}  P[{\bf c}]
\label{eq:marginal}
\end{equation}
for all data points $\mu$. 

\subsection{The SCAP equations}

In the limit $\beta \to \infty$, Eq.~(\ref{partition}) becomes concentrated to the
true cost minima. Looking at Eq.~(\ref{energy}) it becomes obvious that 
$p$ has to scale as $p \propto e^{-\beta {\tilde p}}$ in order to have
some non-trivial effect. In this limit, we find the SCAP equations
(with $\mu\neq\nu$, see Sec.~\ref{methods}):
\begin{eqnarray}
r(\mu,\nu) &= & S(\mu,\nu) - {\max}_{\lambda \ne \nu}\big[S(\mu,\lambda) 
+ a(\mu,\lambda)\big] \nonumber \\ 
r(\mu,\mu) &= & {\max}\big[\!-\tilde p, S(\mu,\mu) 
- {\max}_{\lambda \ne \mu}\{S(\mu,\lambda) 
+ a(\mu,\lambda)\} \big] \nonumber \\ 
a(\mu,\nu) &=& {\min}\Big[ 0,\ r(\nu,\nu)
+\sum_{\lambda \ne \mu} {\max}(0,\ r(\lambda,\nu) \Big]
\label{WP} \\
a(\mu,\mu) &=&
{\min}\Big[{\tilde p},\ \sum_{\lambda \ne \mu} 
{\max}\{0,r(\lambda,\mu)\} \Big] \ . 
\nonumber
\end{eqnarray}
These $2N^2$ equations are closed and can be solved iteratively. Following 
Eq.~(\ref{eq:marginal}) the exemplar $c^*_\mu$ of any data point $\mu$ can be computed 
by maximizing the marginal {\it a posteriori} probability:
\begin{equation}
  c^*_\mu = {\rm argmax}_\nu \big[\, a(\mu,\nu) + r(\mu,\nu)\, \big]
  \label{cstar}
\end{equation}
Compared to original AP, SCAP amounts to an additional 
threshold on the self-availabilities $a(\mu,\mu)$ and the self-responsibilities
$r(\mu,\mu)$: For small enough ${\tilde p}$, $a(\mu,\mu)
\to {\tilde p}$ in many cases, up to  $p = 1$ (or ${\tilde p} = 0)$,
where every site chooses its best first neighbor as its
exemplar. At the same
time, beyond a certain threshold the self responsibility
$r(\mu,\mu)$ is substituted with $-{\tilde p}$. For ${\tilde
p} \to \infty$ ({\it i.e.} $p=0$) the original AP equations are recovered.
 
In practice, this means that variables are discouraged to be self
exemplars beyond a give threshold, even in the case someone is
already pointing at them. The resulting clustering is more stable
and obviously allows for a hierarchical structure where $\lambda$ can point
to $\mu$ that can point to $\nu$ etc. Also loops are possible.
In most of the tests performed (both on artificial and biological cancer
data) clusters are almost tree-like besides a dimer.

%As we change $\tilde p$ we go from the naive affinity measure to the
%greedy limit of the Frey {\it et al.} algorithm. $\tilde p$ plays the role of
%keeping a check on the greediness introduced. Also, at the same time,
%via $\tilde p$ we get a better view of the hierarchal
%structure of the clusters. This gives a clear idea of how the
%clusters were formed and insight on which patterns are wrongly
%classified and why. Such  information is completely lost in the
%limit $\tilde p\to \infty$.

\subsection{Efficient implementation of the algorithm}

The iterative solution of Eqs.~(\ref{WP}) can be implemented in the 
following way:
\begin{enumerate}
\item Define the similarity $S(\mu,\nu)$ for  each set of data points. Choose 
the values of the self-similarity $\sigma$ and of the constraint strength $\tilde p$. 
Initialize all $a(\mu,\nu)=r(\mu,\nu)=0$ 
\item For all $\mu\in\{1,...,N\}$, first update the $N$ {\it responsibilities}
$r(\mu,\nu)$ and then the $N$ {\it availabilities} $a(\nu,\mu)$, using 
Eqs.~(\ref{WP}).
\item Identify the exemplars $c_\mu$ by looking at the maximum value of
$r(\mu,\nu)+a(\mu,\nu)$ for given $\mu$, according to Eq.~(\ref{cstar}).
\item Repeat steps 2-3 till there is no change in exemplars 
for a large number of iterations (we used 10-100 iterations). If not
converged after $T_{max}$ iterations (typically 100-1000), stop the
algorithm.
\end{enumerate}

Three notes are necessary at this point:
\begin{itemize}
\item Step 3 is formulated as a sequential update: For each data point $\mu$, 
all outgoing responsibilities and then all incoming availabilities are updated 
before moving to the next data point. In numerical experiments this was found
to converge faster and in a larger parameter range than the damped parallel
update suggested by Frey and Dueck in \cite{Frey07}. Dependence of the result
on initial conditions was not observed. \\[-0.4cm]
\item The naive implementation of the update equations (\ref{WP}) requires $2N^2$
updates, each one of computational complexity $O(N)$. A factor $N$ can be gained
by first computing the unrestricted max and sum once for a given $\mu$, and then 
implying the restriction only inside the internal loop over $\nu$. Like this, the
total complexity of a global update is $O(N^2)$ and thus feasible even for very 
large data sets.\\[-0.4cm]
\item Belief propagation on loopy graphs is not guaranteed to converge. We observe,
however, efficient convergence of the sequential update over wide parameter ranges.
To handle the possibility of non-convergence, we have introduced a cutoff in the
number of iterations. If this is reached, the algorithm stops, and the actual
parameter combination is discarded.
\end{itemize}
 
\subsection{Extracting cluster signatures}

In many clustering tasks input data consist of high-dimensional vectors, a specific 
example being genome-wide microarrays. Frequently only few components of these 
vectors carry useful information about the cluster structure, extracting such cluster
signatures is of crucial importance in understanding the mechanisms behind the 
cluster structure. 

In the following, we will use the specific case of microarray data. Therefore we
use the notion gene for a component of the input vector, even if at this stage the
discussion is still general. The total number of genes is denoted by $G$.
We propose a simple measure of the influence of single genes on the total similarity 
measure of a cluster, as compared to random choices of the exemplar selection ${\bf c}$. 
For simplicity, we assume the similarity between data points 
$x_\mu = (x_\mu^1,...,x_\mu^G)$ and $x_\nu = (x_\nu^1,...,x_\nu^G)$ to be
additive in single-gene contributions
\begin{equation}
S(\mu,\nu) = \sum_{i=1}^G s(x_\mu^i,x_\nu^i)\ .
\end{equation}
This is true, {\it e.g.}, for the Pearson correlation or the negative square
Euclidean distance. It can be easily generalized to similarity measures 
which are given by a monotonous function of a sum over gene contributions (like 
the negative of the Euclidean distance which is the square root of the sum of
single-gene contributions).

Having found a clustering given by the exemplar selection ${\bf c}$, we can 
calculate the similarity of a cluster $C$ defined as a connected component
of the directed graph given by ${\bf c}$. It is given by
\begin{equation}
S(C) = \sum_{i=1}^G S^i(C)\ ,
\end{equation}
as a sum over single-gene contributions
\begin{equation}
S^i(C) =  \sum_{\mu\in C} s(x_\mu^i,x_{c_\mu}^i)
\end{equation}
These have to be compared to random exemplar choices which are characterized by
their mean
\begin{equation}
S_0^i(C) = \sum_{\mu\in C} \frac 1N \sum_{\nu=1}^N  s(x_\mu^i,x_\nu^i)
\end{equation}
and variance
\begin{equation}
[\Delta S_0^i(C)]^2 = 
\sum_{\mu\in C} \left\{ \frac 1N \sum_{\nu=1}^N  s(x_\mu^i,x_\nu^i )^2
- \left[ \frac 1N \sum_{\nu=1}^N  s(x_\mu^i,x_\nu^i )  \right]^2 \right\}
\end{equation}
The relevance of a gene can now be ranked according to 
\begin{equation}
I_i(C) = \frac{S^i(C)-S_0^i(C)}{\Delta S_0^i(C)}
\end{equation}
which measures the distance of the actual $S^i(C)$ from the distribution of
random exemplar mappings. Genes can be ranked according to the value of 
$I_i(C)$, highest-ranking genes are considered a cluster signature. The same 
procedure can be carried through for each cluster independently, but also for 
cluster combinations.

\section{Application to biological data}

\subsection{Iris data}

The data consist of measurements of sepal length, sepal width, petal
length and petal width, performed for 150 flowers, chosen from  three
species of the flower Iris. It is a benchmark problem for clustering
\cite{Duda73}. Super-paramagnetic clustering is able to cluster 125 of 
the data points correctly, leaving 25 points unclustered \cite{blatt96}.

When we apply AP on Iris data, we identify
three clusters making 16 errors. With SCAP, we identify them with 
just nine errors. We use the Manhattan distance measure for
the similarity function, i.e $S(\mu,\nu)=-\sum_{i=1}^{4}
|x_{\mu}^i-x_{\nu}^i|$.

We saw that the species Iris Setosa separates without any 
errors. On increasing the value of $\tilde p$, the Iris Setosa cluster
stays intact and the clusters for Versicolor and Virginica merge
with each other,  reflecting the fact
that they are closer to each other than to Setosa. The errors
occur because some samples from these species were closer to samples
from other species than to their own.

\subsection{Brain cancer data}
We used a test data set monitoring the expression levels of more than 7000 genes 
for 42  patients, which were previously correctly classified into  5 diagnosis types by 
an {\it a posteriori} assessment method \cite{Pomeroy02} (10 medulloblastoma,
10 malignant glioma, 10 atypical teratoid/rhabdoid tumors, 4 normal  cerebella, 8 
primitive neuroectodermal tumors). Each array was filtered, log-normalized 
to mean zero and variance one, resulting in $G=6010$ genes. Due to this choice Pearson
correlation and negative square Euclidean distance are equivalent. The diagnosis 
information was not used during clustering, but only for checking the algorithmic outcome.

{\it Imposing five clusters in AP and SCAP:}
Since we knew that the correct clustering was to identify five different patterns, 
our first approach was to tune  $\sigma$ and $\tilde p$ in order to get five clusters. 
First, we fixed $\tilde p$ to infinity (original AP) and changed $\sigma$ finding 
around $\sigma \sim 120$ the desired number of 5 clusters with 8 errors. The error was
calculated {\it a posteriori} by counting every data point which referring to an exemplar 
of a different diagnosis. Next we fixed $\sigma$ to a sufficiently large value 
(the result becomes insensitive on $\sigma$ once the latter takes large values),
and we changed  $\tilde p$. In this case, for $\tilde p=12$ we got 6 clusters with 8
errors, for $\tilde p=16$ 4 clusters with again 8 errors. 5 clusters were not found
to correspond to any extended $\tilde p$-region. Note that in both cases all 
errors occur in the last cluster: samples supposed to take diagnosis 5  (PNET) 
rarely find an exemplar of the same class. Instead they distribute over the other 
four diagnoses.

\begin{figure}[!tpb]%figure1
\centerline{\includegraphics[width=7 cm]{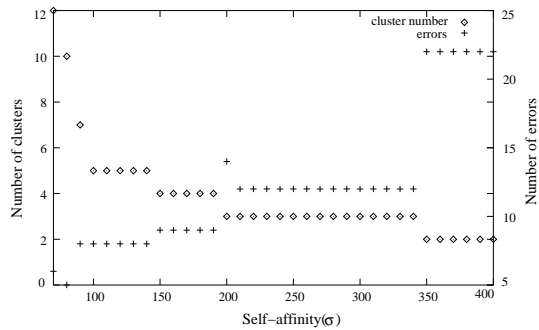}}
\caption{Plot of number of clusters as a function of self-affinity 
$S(\mu,\mu)\equiv-\sigma$, for $\tilde p = \infty$ (the original 
AP algorithm). Based on this we would conclude that the data has three 
nontrivial clusters.}
\label{fig:01}
\end{figure}

{\it Clustering with AP:}
Then, instead of fixing the number of clusters, we changed $\sigma$ continuously 
for $\tilde p\to\infty$. We counted the number of clusters and of errors as a 
function of $\sigma$, see Fig.~\ref{fig:01}. The algorithm ground state 
(configuration of maximum marginals values) in the limit of $\sigma\to\infty$ 
is a single cluster. 

The first non trivial clustering occurs when the number of clusters remain 
unchanged for a stable range of $\sigma$ values. In this preliminary study, we 
took that to be the actual predicted data clustering. Hence, by looking at 
Fig.~\ref{fig:01}, we would conclude that there are three well-distinguishable 
clusters in the present data set. Look, however, to the number of errors: It is
found to be 14-15 in this range, basically due to the wrong assignment of two
entire classes to only three exemplars. Four or five clusters can be imposed and 
lead to lower error values, but require fine-tuning of $\sigma$.

\begin{figure}[!tpb]%figure2
\centerline{\includegraphics[width=7 cm]{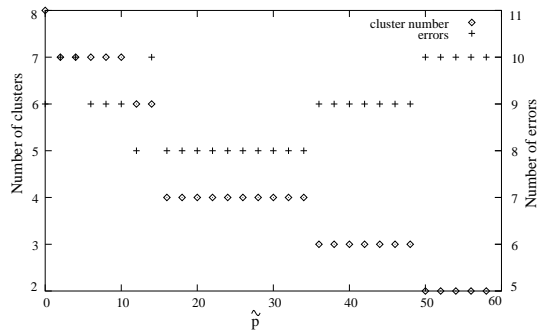}}
\caption{Plot of number of cluster and errors of SCAP as a function of cutoff 
parameter $\tilde p$. This plot suggests that the data has four clusters.}
\label{fig:02}
\end{figure}

{\it Clustering with SCAP:}
We than fix $\sigma$ to be very large and change only $\tilde p$. For $\tilde p=0$
we start with seven clusters, this number decreases rapidly as $\tilde p$
increases, see Fig.~\ref{fig:02}. As before, the point at which the number of 
clusters is robust against changes in $\tilde p$ was taken as the best SCAP 
clustering. From Fig.~\ref{fig:02} we conclude that SCAP identifies $4$ clusters. 
The number of errors in classification is 8. 

%Let us show in more detail how the clusters look as we change the $\tilde p$. 
%Usually the modified AP is able to find either the correct 
%number or a better estimation of the true cluster number in this way.

\begin{figure*}
\begin{center}
\includegraphics[width=17.5cm]{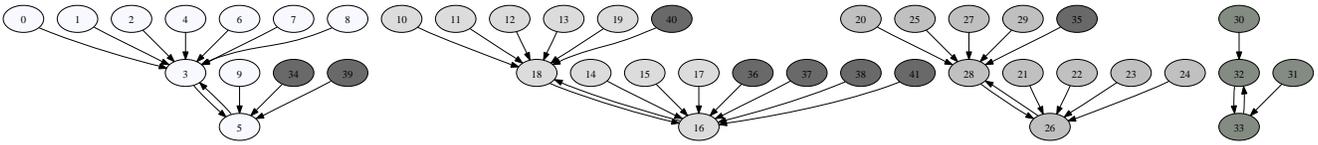}
\end{center}
\caption{Graphical presentation of clusters at $\tilde p=30$. Increasing 
$\tilde p$ did not separate the data set (10-19) and (34-41). Instead they 
together make one big cluster.} 
\label{fig:05}
\end{figure*}

Right from $\tilde p=0$, where each data point chooses its closest neighbor as 
his exemplar, errors are due to misclassifications of the fifth diagnosis (PNET).
The other data points select exemplars of the same diagnosis, but various clusters
of same diagnosis exist. Only in the case of four clusters, as shown in 
Fig.~\ref{fig:05}, each of the first four diagnoses is assembled in an isolated
cluster, with the PNET arrays distributed over the three cancer-related clusters.
The normal tissue (30-33 in the figure) is well-separated from all others. Only
if we go towards three clusters, it merges with diagnosis type (0-9) 
(medulloblastoma), showing that these two are closer in expression in between them
than compared to others. A more detailed analysis of the brain cancer data is 
provided in the supplementary material.

%\subsubsection{Structure of cluster at different $\tilde p$}
%At $\tilde p=0$, the constraint is completely relaxed and every data point chooses 
%its first neighbor  as its exemplar ({\it i.e.} the most similar one to the data 
%point under consideration according to the function $S$). Look at Figs.~\ref{fig:03}
%to \ref{fig:07} for the change in the clustering number and structure as we 
%increase $\tilde p$. The grey scale of the nodes represents different diagnosis. 
%Data points between (0-9), (10-19), (20-29), (30-33) and (34-41) had same diagnosis. 
%Clearly right from $\tilde p=0$, we see that the last set (34-41) is mixed up with 
%other clusters. At $\tilde p =5$, the data clusters into five groups. Here,  
%groups (0-9), (20-29) and (30-33) have already merged, while (10-19) is still 
%split into two clusters. 
%On the other hand, the points which should have been included in the last cluster 
%(34-41) are distributed among all the others. For $\tilde p=8$, we 
%see four clusters, errors occurring only in the assignment of points 
%$34-41$. Most of them seem to belong to group (10-19), 
%suggesting that they have similar expressions. Our conclusion of the previous section, 
%that the data set has only 4 distinguishable clusters, is coherent with this alternative 
%presentation. As we change $\tilde p$, we can get a possible hierarchal picture 
%of the clusters forming process. As we go to even higher $\tilde p$, 
%we see that groups (0-9) and (30-33) merge, implying that these
%two are closer together than the rest.

Note that SCAP also provides information about the internal organization inside
the clusters. We find, {\it e.g.}, that the misclassified patterns are always
peripherical cluster elements. No other data point refers to them. This information
is lost in AP. 
%We show in Fig.~\ref{fig:08} the graphical structure of the 
%data-exemplar mapping for original AP. 
Due to the hard constraint all points belonging
to the same cluster refer to the same exemplar, and information about the internal 
cluster structure is not contained in ${\bf c}^\star$. A graphical representation of
the cluster structure in this case is contained in the supplementary material.

In \cite{Pomeroy02} data were clusterized using hierarchical clustering. Even if the 
overall cluster structure is similar to the one we found, there is no clear-cut 
clustering into 4-5 classes, some arrays (well-clustered with SCAP) were only added at very 
late stages of hierarchical clustering. The global nature of SCAP leads to a better
clustering performance than the local and greedy hierarchical clustering.

Another interesting point comes from the comparison of our clustering results with the
supervised classification results of \cite{Dettling04}. There, a number of state-of-the-art
classification algorithms is applied, with training sets containing 2/3, test sets 
containing 1/3 of the data points. Dettling finds that the minimal generalization error
made is 23.8\%, corresponding to ca. 10 errors on a data set of cardinality 42. It is
interesting to note that SCAP in the clustering corresponding to 4 clusters makes only
8 errors. Note that training in \cite{Dettling04} is done on a subset of patterns, but 
supervision in this case seems to add no valuable information to the unsupervised
clustering results.

Last but not least, we use the procedure described above to extract cluster signatures
in the most stable case of four clusters depicted in Fig.~\ref{fig:05}. The lists
of the highest ranking genes together with their relevance value $I_i(C)$ is given
in the supplementary material. The number of statistically relavant genes (we consider 
a threshold $I_i(C)\simeq 3$) depends on the cluster and is largest for the normal tissue 
(42 genes), it is much smaller in particular for the first cluster (ca. 4 genes). If we
take the first 15-25 genes per cluster, {\it i.e.}, an overall signature of 60-100
genes, we already find basically the same clustering as before, only two new errors
of previously well-assigned patterns appear. At gene signatures 120-240, only one
of these errors survives. We therefore find that the signature found in this way carries
most of the information needed for the clustering. Note also, that due to the fact that
in an unsupervised way we did not separate the fifth diagnosis type into a single
cluster, we do not have by definition a cluster signature for this cancer type.

%\begin{figure*}
%\begin{center}
%\includegraphics[width=17cm]{brp0.ps}
%\end{center}
%\caption{Graphical presentation of clusters at $\tilde p=0$. At this value of 
%$\tilde p$, data points refer to their nearest neighbor. One can already see 
%that some points corresponding to last diagnosis (points 34-41) are very close to 
%points 10-19.} 
%\label{fig:03}
%\end{figure*}

%\begin{figure*}
%\begin{center}
%\includegraphics[width=17.5cm]{brp5.ps}
%\end{center}
%\caption{Graphical presentation of clusters at $\tilde p=5$. At this point 
%the data set (0-9),(20-29) and (30-33) are already clustered together} 
%\label{fig:04}
%\end{figure*}

%\begin{figure*}
%\begin{center}
%\includegraphics[width=17.5cm]{brp10.ps}
%\end{center}
%\caption{Graphical presentation of clusters at $\tilde p=10$. Though still 
%we have four clusters, second cluster from the left has compactified.} 
%\label{fig:06}
%\end{figure*}

%\begin{figure*}
%\begin{center}
%\includegraphics[width=17.5cm]{brp15.ps}
%\end{center}
%\caption{Graphical presentation of clusters at $\tilde p=15$. At this point, 
%the clusters (0-9) and (30-33) have merged together, suggesting a closeness among 
%them relative to other clusters.} 
%\label{fig:07}
%\end{figure*}

%\begin{figure*}
%\begin{center}
%\includegraphics[width=17.5cm]{brpinftymu70.ps}
%\end{center}
%\caption{Graphical presentation of clusters in the limit of 
%$\tilde p \to \infty$. Here, unlike the above figures, all the information 
%about the hierarchical structure is  lost. Hence it is almost impossible 
%to deduce the sources of error from this clustering.}
%\label{fig:08}
%\end{figure*}

\subsection{Other benchmark cancer data}

{\it Lymphoma cancer data:}
We used a data set of 62 patients for 4026 genes, showing 3 different diagnosis
\cite{Alizadeh00}.
In the limit of $\tilde p$ going to infinity, we find the first nontrivial
clustering for $\sigma$ between $150-250$. In this regime AP group data 
into 3 sets, making with 3 error. For very high $\sigma$ and 
varying $\tilde p$, the 3-groups clustering becomes more stable and robust,
while  the algorithm makes just one assignment prediction error.
In this case, Dettling finds a minimal generalization error of 0.95\%, corresponding
to less than one error in 62 patterns. Supervision adds some information, even if
clustering itself makes only one error.

{\it SRBCT cancer data:}
This set has 63 samples with 2308 genes and 4 expression diagnosis patterns
\cite{Kahn01}.
For $\tilde p$ going to infinity, the best tuning-robust estimates groups 
cluster data into 5 clusters making as many as 22 errors. On the other hand, 
with finite $\tilde p$, SCAP finds a regime of 4 clusters, making 
only 7 assignment errors.
Here, Dettling reports only 1.24\% generalization error in supervised 
classification, corresponding to less than one error on 63 patterns. Classification
thus performs considerably better than clustering alone.

{\it Leukemia:}
This set has 72 samples with 3571 genes and 2 diagnoses \cite{Golub99}. 
In the case of 
infinite $\tilde p$, the original AP groups data into 2 clusters with 
4 errors, while for variable $\tilde p$ (fixing $\sigma$ very large) modified AP 
finds 2 clusters with 2 errors.
Also classification leads to 2.5\% of errors, a result which is slightly better 
than our clustering result.

%\begin{methods}

\section{Methods}
\label{methods}

In the process of choosing exemplars, we need to calculate marginals
\begin{equation}
P_\mu(c_\mu) = \sum_{\{c_\nu \}_{\nu \ne \mu}} P({\bf c})
\label{marginal}
\end{equation}
$P_\mu(c)$ is the probability that data point $\mu$ chooses point $c$ as its exemplar.
The calculation of marginals can be done iteratively via a message-passing
algorithm called Belief Propagation (BP) \cite{Yedidia05}. It is exact on tree factor graphs 
but usable heuristically in the general case. Together with a generalized larger family of 
message-passing algorithms, it was shown to be very 
powerful in solving NP-hard combinatorial problems on locally tree-like structures 
\cite{Mezard02,Hartmann05}. Recently, the applicability of BP 
was also shown to be efficient in some important problems giving rise to dense and 
loopy factor graphs \cite{Kabashima,Alfredo-Riccardo}. 

\begin{figure}[!tpb]
\centerline{\includegraphics[width=7cm]{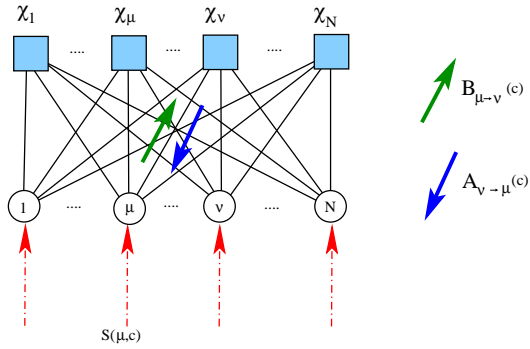}}
\caption{SCAP factor graph and direction of sent messages. Variables are represented by circles,
constraints by squares.}
\label{AP-FG}
\end{figure}

\begin{figure}[!tpb]
\centerline{\includegraphics[width=7cm]{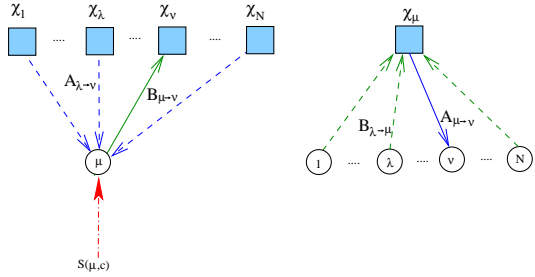}}
\caption{Pictorial representation of messages flow.}
\label{messages}
\end{figure}

Looking at figures (\ref{AP-FG}) and (\ref{messages}), BP computes beliefs $P^{BP}_\mu(c)$ 
for the marginal probabilities  as products of messages $A_{\lambda \to \mu}(c)$ coming 
from each compatibility constraint, times the local prior computed 
as the exponential of the similarity
between point $\mu$ and its putative exemplar $c$. Up to overall normalization, we write: 
\begin{equation}
P^{BP}_\mu(c) \propto \prod_\lambda A_{\lambda \to \mu}(c) e^{\beta S(\mu,c)}
\label{eq:mar}
\end{equation}
where $\beta$ plays the role of an annealing parameter measuring the relative
importance given to the priors compared to the information passed by the messages.
Message $A_{\lambda \to \mu}(c)$ can be interpreted, as the probability that
constraint $\lambda$ alone forces $\mu$ to select exemplar $c$. It can be calculated 
via the following self-consistent equations
\begin{eqnarray}
A_{\mu \to \nu}(c) &\propto& 
\sum_{\{c_\lambda\}}
\prod_{\lambda \ne \nu} 
B_{\lambda \to \mu}(c_\lambda)
\chi^{(p)}_\mu 
(\{c_\lambda \},c) 
\label{firstBP-A} \\
B_{\mu \to \nu}(c) &\propto& 
\prod_{\lambda  \ne \nu}  A_{\lambda \to \mu}(c) \exp (\beta S(\mu, c)) 
\label{firstBP-B}
\end{eqnarray}
where the $N^2$ functions $B_{\mu \to \nu}(c)$ can be seen as probabilities that data point 
$\mu$ chose $c$ to be its exemplar if constraint $\lambda$ were absent in Eq.~(\ref{partition}). 
These probabilities are called {\it cavity probabilities}, because the disregarding of one 
data point / constraint effectively carves a cavity in the original factor graph. 
The {\it link direction} of functions $A$s and $B$s is shown if fig.~(\ref{AP-FG}) 
together with the problem's factor graph. Fig.~(\ref{messages}) shows a 
pictorial representation of the flow of messages (\ref{firstBP-A}) and (\ref{firstBP-B}).

Along the lines of \cite{Frey07} and \cite{Frey07-B}, but bearing in mind the
modified form for the compatibility constraints, eq.~(\ref{firstBP-A}) can be simplified 
after a few manipulations in the following way, depending on cases:
\begin{eqnarray}
(\mu = \nu) \; \wedge \; (c = \mu) &\to& A_{\mu \to \mu}(\mu) = \frac{1}{Z^A_{\mu \to \mu}}
\prod_{\lambda \ne \nu} 
\sum_{\{c_\lambda\}}
B_{\lambda \to \mu}(c_\lambda) \nonumber \\
&=& \frac{1}{Z^A_{\mu \to \nu}} \nonumber  \\
(\mu = \nu) \; \wedge \; (c \ne \mu) &\to& A_{\mu \to \mu}(c : c\ne \mu) 
= \frac{1}{Z^A_{\mu \to \mu}}
\big[ 
p + \nonumber \\
&& (1-p) \prod_{\lambda \ne \nu} (1 - B_{\lambda \to \mu}(\mu))
\big]  \label{AA} \\
(\mu \ne \nu) \; \wedge \; (c = \mu) &\to& A_{\mu \to \nu}(\mu) 
= \frac{1}{Z^A_{\mu \to \nu}}
\left[ 
p + (1-p) B_{\mu \to \mu}(\mu))
\right]  \nonumber \\
(\mu \ne \nu) \; \wedge \; (c \ne \mu) &\to& A_{\mu \to \nu}(c : c \ne \mu) 
= \frac{1}{Z^A_{\mu \to \nu}}
\big[ 
p + (1-p)* \nonumber \\
&& * \big(( B_{\mu \to \mu}(\mu)) + \prod_{\lambda \ne \nu} 
(1 - B_{\lambda \to \mu}(\mu)) \big)
\big] \nonumber
\end{eqnarray}
with $Z^A_{\mu \to \nu}$ being normalization constants.
%\begin{eqnarray}
%Z^A_{\mu \to \mu} &=& 1 + (N-1) \big[ 
%p + (1-p) \prod_{\lambda \ne \nu} (1 - B_{\lambda \to \mu}(\mu))
%\big] \\
%Z^A_{\mu \to \nu} &=& p + (1-p) B_{\mu \to \mu}(\mu)) + \\
%& &  (N-1) \big[ 
%p + (1-p) ( B_{\mu \to \mu}(\mu)) + \prod_{\lambda \ne \nu} 
%(1 - B_{\lambda \to \mu}(\mu)) )
%\big] \nonumber
%\end{eqnarray}

It is remarkable that the number of effectively independent quantities present in 
Eqs.~(\ref{firstBP-A}) is much smaller than the apparent $O(N^3)$ real valued
numbers. Indeed, functional messages $A$ take only 2 
different values: $A_{\mu \to \nu}(\mu)$, that from now on will be called  $A(\nu, \mu)$ 
to avoid index redundancy, and  $A_{\mu \to \nu}(c : c \ne \mu) = {\hat A}_{\mu \to \nu}$ 
independently on $c$, as long as it is $\ne \mu$. The exchange of indexes 
in $A(\mu,\nu)$ is pure convention, but it has been introduced for coherency with the 
definition of availabilities given in \cite{Frey07}.
It follows immediately from the normalization condition that
${\hat A}_{\mu \to \nu} = (1 - A(\nu, \mu))/(N-1)$.

For the cavity probability functions, manipulation of Eq.~(\ref{firstBP-B}) involving the use 
of the last normalization condition and of the normalization constant rescaling, leads to
\begin{eqnarray}
(\mu = \nu) \; \wedge \; (c = \mu) &\to& B_{\mu \to \mu}(\mu) = R(\mu,\mu) 
= \frac{e^{\beta S(\mu,\mu)}}{Z^B_{\mu \to \mu}} \nonumber \\
(\mu = \nu) \; \wedge \; (c \ne \mu) &\to& B_{\mu \to \mu}(c : c \ne \mu) = \nonumber \\
& & = \frac{1}{Z^B_{\mu \to \mu}}
\frac{(N-1) A(\mu,c) e^{\beta S(\mu,c)}}{1 - A(\mu,c)}  \nonumber \\
\end{eqnarray}
\begin{eqnarray}
(\mu \ne \nu) \; \wedge \; (c = \mu) &\to& B_{\mu \to \nu}(\nu) = R(\mu,\nu) 
= \frac{e^{\beta S(\mu,\nu)}}{Z^B_{\mu \to \nu}} 
\label{BB} \\
(\mu \ne \nu) \; \wedge \; (c \ne \nu) &\to& B_{\mu \to \nu}(c : c \ne \nu) = \nonumber \\
& & = \frac{1}{Z^B_{\mu \to \nu}}
\frac{(N-1) A(\mu,c) e^{\beta S(\mu,c)}}{1 - A(\mu,c)} 
\nonumber
\end{eqnarray}
with $Z^B_{\mu \to \nu}$ guaranteeing  normalization $\sum_c B_{\mu \to \nu}(c)=1$.
%\begin{equation}
%Z^B_{\mu \to \nu} = e^{\beta S(\mu,\nu)} + (N-1) \sum_{c : c \ne \nu} 
%\frac{A(\mu,c) e^{\beta S(\mu,c)}}
%{1 - A(\mu,c)}
%\label{norm-B}
%\end{equation} 
Messages  $B_{\mu \to \nu}(\nu) \to R(\mu, \nu)$ have been also renamed with a 
symbol coherent with 
the {\it responsibility-availability} notation of \cite{Frey07}. It can be seen that self 
consistent equations close into the ${\it O}(N^2)$ quantities $A(\nu, \mu)$ and 
$R(\mu, \nu)$ alone. Indeed, the effective dependence on the exemplar choice is dropped, 
and the computational size of the problem reduces by a factor $N$.
The set of equations of $A$ and $R$ can be solved iteratively via the BP algorithm. 
The case in which one is interested not in the whole form of the posterior 
probability function, but only in retaining information about the most probable exemplar
chosen by each data point, can be seen to be equivalent to taking the $\beta \to \infty$ limit 
where availabilities $a$ and responsibilities $r$ are introduced in the following way:
\begin{eqnarray}
A(\mu,\nu) &=& e^{\beta a(\mu,\nu)} e^{\beta {\hat a}_{\nu \to \mu} }\\
{\hat A}_{\nu \to \mu} &=& e^{\beta {\hat a}_{\nu \to \mu} }\\
R(\mu,\nu) &=& e^{\beta r(\mu,\nu) + \beta max_{c : c \ne \nu} \{ r_{\mu \to \nu}(c : c \ne \nu) \} } \\
B_{\mu \to \nu}(c : c \ne \nu) &=& e^{\beta r_{\mu \to \nu}(c : c \ne \nu)}
\end{eqnarray}
and treating the exponential scaling in a regime where prior similarities between data points $S$ 
are of the same order of magnitude of the valued of $a$ and $r$. The rescaling of responsibilities 
can be freely done as it does not change the number of independent variables. From the last definitions
one is led to equations
\begin{eqnarray}
A(\mu,\nu) &=& \frac{1}{1 + (N-1) e^{-\beta a(\mu,\nu) }}  
\label{ABETA} \\
R(\mu,\nu) &\sim& \frac{1}{1 + e^{-\beta r(\mu,\nu) } }
\label{BBETA}
\end{eqnarray}
where the last relation already assumes a large-$\beta$ limit with non-degeneracy 
of the most probable value of the cavity probabilities. This hypothesis is equivalent to 
having non-degenerate choices of exemplars for all data points, i.e., to
the existence of a single optimal clustering identified via the SCAP algorithm as the unique 
ground state of the system energy (\ref{energy}).
This is a sensible assumption, but it is not always satisfied in interesting cases. Studying 
the degenerate number and behavior of clustering choices is another crucial question that is 
only partially answered by
the introduction of the relaxation parameter $p$ and will be the subject of further work beyond 
this paper. In the large $\beta$ regime, in order to work with quantities all with the same 
scaling, it is useful to define 
\begin{equation}
p = e^{-\beta {\tilde p}}
\label{ttbeta}
\end{equation}
and consider ${\tilde p}$, $a$ and $r$ fixed varying $\beta$. 
Equating Eqs.~(\ref{ABETA}) and (\ref{BBETA}) with Eqs.~(\ref{AA}) and (\ref{BB}) respectively, 
and extracting the leading terms in the large $\beta$ limit assuming no degeneracy, the following 
equations are found, using Eq.~(\ref{ttbeta}):
\begin{eqnarray}
r(\mu,\nu) &= & S(\mu,\nu) - {\max}_{\lambda \ne \nu}\big[S(\mu,\lambda) 
+ a(\mu,\lambda)\big] \nonumber \\ 
r(\mu,\mu) &= & S(\mu,\mu) 
- {\max}_{\lambda \ne \mu}\big[S(\mu,\lambda) 
+ a(\mu,\lambda)\big] \\ 
a(\mu,\nu) &=& {\min}\Big[ 0,{\max}(-{\tilde p}, {\min}(0,\ r(\nu,\nu)))
+\sum_{\lambda \ne \mu} {\max}(0,\ r(\lambda,\nu) \Big]
\nonumber \\
a(\mu,\mu) &=&
{\min}\Big[{\tilde p},\ \sum_{\lambda \ne \mu} 
{\max}\{0,r(\lambda,\mu)\} \Big] \ . 
\nonumber
\label{WP-0}
\end{eqnarray}
Making another change of variables redefining the self-responsibilities as 
\begin{equation}
r'(\mu,\mu) = {\max}\big[ -{\tilde p}, r(\mu,\mu) \big]  - {\max}_{\mu}\{ r(\mu,\mu) \} \to r(\mu,\mu)
\end{equation}
we get, in terms of the rescaled quantities,
\begin{equation}
{\max}(-{\tilde p}, {\min}(0,\ r(\mu,\mu))) = r(\mu,\mu)
\end{equation}
leading to SCAP equations (\ref{WP})
\footnote{Note that the last change of variable via the subtraction 
of the overall quantity ${\max}_{\mu}\{ r(\mu,\mu) \}$ is redundant if self-responsibilities are negative, 
as it is usually the case.}.
After convergence, marginals can be written \cite{Yedidia05, Frey07} as
\begin{eqnarray}
P_{\mu}(c) &\sim& P^{BP}_{\mu}(c) \propto \prod_\lambda A_{\lambda \to \mu}(c) \exp (\beta S(\mu,c)) 
\nonumber \\
&\propto& R(\mu,c) A(\mu,c) \propto e^{\beta (a(\mu,c) + r(\mu,c))}
\label{marginals}
\end{eqnarray}
In the $\beta \to \infty$ limit, one can write equation~ (\ref{cstar}).

%\end{methods}

\section{Discussion}

Affinity Propagation is a new powerful tool for unsupervised clustering. It has many
very strong points. First it is very efficient, convergence to the final clustering
is very fast, the latter appears to be independent on the initialization of messages.
Second, due to its hard constraints AP identifies exemplars which are prototypical
data points representing a whole cluster. 

This last point is, however, also a first limitation of the original AP algorithm.
If clusters cannot be well-represented by a single cluster exemplar, AP has to fail.
The hard constraint renders the algorithm greedy, and small fluctuations in the
similarity measure may trigger avalanches in the exemplar choice leading to
different clusterings for only slightly modified model parameters.

We have introduced a soft-constraint version of affinity propagation which is able
to cure a part of these problems without loosing the efficiency of the original AP:
\begin{itemize}

\item By relaxing the hard constraint on clusters exemplars, we could
introduce a parameter ($\tilde p$) controlling 
the algorithm greediness. $\tilde p$ is a better tuning parameter than 
$\sigma$ (it is more informative and leads to more robust and stable 
clustering) and it is easier to interpret the statistical meaning of its 
tuning process.\\[-0.4cm]

\item Clusters are more robust than in the original formulation 
of the algorithm. Moreover, even though a second {\it a priori} free-parameter 
is introduced, the overall dependence of the algorithm on free parameters is 
reduced, and an optimal tuning strategy naturally emerges.\\[-0.4cm]

\item The cluster structure can be efficiently probed. This
concerns the internal structure of the clusters since SCAP is able to
identify central and peripherical nodes of each clusters, as well as
the hierarchical organization leading to a process of cluster merging if 
cluster number is reduced by looking to less fine structures.\\[-0.4cm]

\item In the case of high-dimensional data, the relation between data points 
and their exemplars can be used to extract a sparse cluster signature. In the
case of brain tumors, we have found that 20-40 genes per cluster are sufficient
to reproduce almost the same clustering as found using all genes.\\[-0.4cm]

\end{itemize}

We conclude that SCAP is more efficient than AP in particular in the case
of noisy, irregularly organized data - and thus in biological applications concerning
microarray data. The computational efficiency of SCAP allows there to treat also very
large data sets.

\section*{Acknowledgments}{}
We acknowledge very useful discussions with Alfredo Braunstein, Andrea Pagnani and 
Riccardo Zecchina. M.L. would like to thank the Malawi 
Polytechnic for hospitality during the preparation of the manuscript. The work of S.
and M.W. was supported by the EC via the STREP GENNETEC 
(``Genetic networks: emergence  and complexity'').

\end{document}